\documentclass{article} 

\usepackage{amsmath}
\usepackage{graphicx}
\makeatother
\begin{document}

\title{Robust frequency stabilization of multiple spectroscopy lasers with
large and tunable offset frequencies}


\author{A. Nevsky,$^{1,*}$ S. Alighanbari,$^1$, Q.-F. Chen$^1$, I. Ernsting$^1$, S. Vasilyev$^1$,
\\
 S. Schiller$^1$, G. Barwood$^{2}$, P. Gill$^{2}$, N. Poli$^{3}$, G. Tino$^{3}$
\\
$^1$Institut f\"ur Experimentalphysik, Heinrich-Heine-Universit\"at D\"usseldorf, 
\\
40225 D\"usseldorf, Germany
\\
$^{2}$National Physical Laboratory, Hampton Road, Teddington, Middlesex,
\\
TW11 0LW, UK
\\
$^{3}$Dipartimento di Fisica e Astronomia, Universit\'a di Firenze, 50019 Sesto Fiorentino, 
\\
Italy
\\
$^*$Corresponding author: alexander.nevsky@uni-duesseldorf.de}
\maketitle

\begin{abstract}
We demonstrate a compact and robust device for simultaneous absolute
frequency stabilization of three diode lasers whose carrier frequencies
can be chosen freely relative to the reference. A rigid ULE multi-cavity
block is employed, and, for each laser, the sideband locking technique
is applied. Useful features of the system are a small lock error,
computer control of frequency offset, wide range of frequency offset,
simple construction, and robust operation. One concrete application
is as a stabilization unit for the cooling and trapping lasers of
a neutral-atom lattice clock. The device significantly supports and
improves the operation of the clock. The laser with the most stringent
requirements imposed by this application is stabilized to a linewidth
of 70 Hz, and a residual frequency drift less than 0.5 Hz/s. The carrier
optical frequency can be tuned over 350 MHz while in lock. 
\end{abstract}

\noindent 
In precision spectroscopic experiments with cold atoms and molecules
the need arises to stabilize the frequency of lasers to such a level
that both the residual absolute frequency error and the residual linewidth
of the laser radiation are well below the linewidth of the transition
of the species under study. For example, in the application of the
strontium optical lattice clock, one cooling laser should have its
absolute frequency stabilized and its linewidth reduced to a level
of 1 kHz \cite{Poli2006}.

Often, a frequency reference capable of providing adequate frequency
control is not available at the required optical frequency for the
spectroscopic application. For example, the frequency reference may
be a rigid and therefore non-tunable ULE cavity, or a ro-vibrational
transition of a molecular gas in a cell. The spacing between neighboring
frequency markers (TEM$_{00}$ modes or molecular transitions), $\Delta\omega_{f}/2\pi$,
will typically be on the order of 1 to a few GHz, and therefore the
offset between the goal frequency and the reference frequency may
have to be as large as  $\sim1$ GHz. Such a gap can in principle
be bridged with one or more acousto-optic modulators. 
Another approach is the serrodyne technique recently reported \cite{Houtz2010},
in which a single sideband is produced by means of waveguide electro-optic phase modulator (EOPM). 

Here, we utilize the sideband locking technique \cite{Thorpe2008}.
It is based on the use of a EOPM, driven by a simple, easily available
electrical waveform, a phase-modulated (PM) harmonic signal (Alternatively,
an amplitude-modulated harmonic signal could be used). This approach
has the advantage of being fiber-based and compact, and of allowing
bridging gaps of many GHz, limited only by the EOPM bandwidth. It
also provides an unambiguous lock state, and introduces (for most
purposes) a negligible lock frequency offset error. Moreover, it was
previously shown that the lock instability may reach a level comparable
to that of a standard lock using the carrier only \cite{Thorpe2008}.

We apply this technique for the stabilization of three diode lasers
(689 nm, 813 nm, 922 nm) to a compact reference cavity unit. This
laser set is required for cooling and trapping strontium atoms in
an optical lattice, and we thus realize a frequency stabilization
system (FSS) which thanks to its compactness may serve as a prototype
for transportable or space optical clocks \cite{SOC2}.

\begin{figure}
\includegraphics[width=7.5cm]{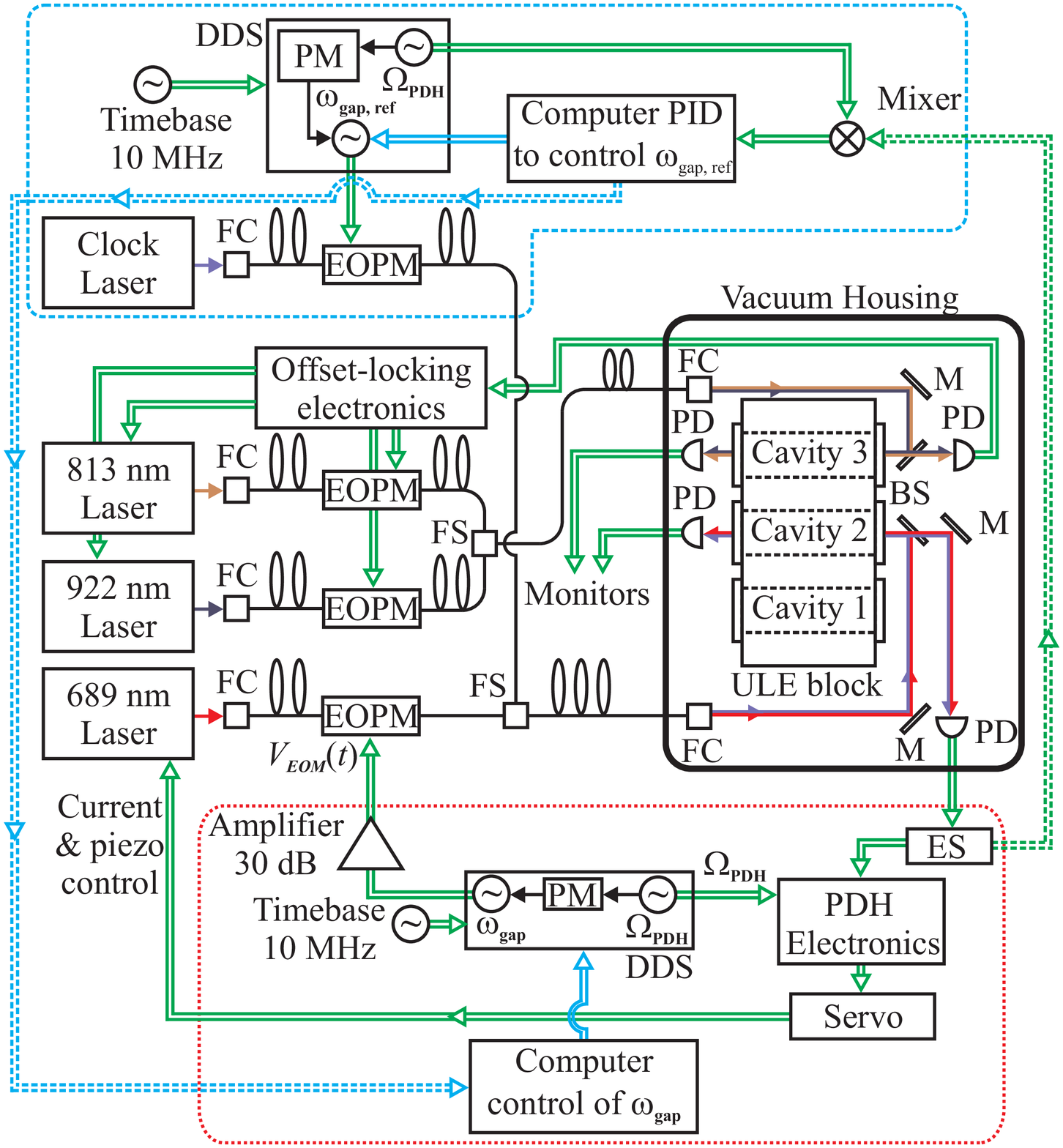}
\caption{Schematic of the FSS. The offset-locking electronics for the 813~nm
and 922~nm lasers are the same as what is indicated by the red dotted
box. FC: fiber collimator, FS:
fiber splitter, ES: electronic splitter, BS: beam splitter, PD: photodetector,
M: mirror, PM: phase modulation unit. DDS: direct digital synthesizer.\label{fig:Principle}}
\end{figure}

\begin{figure}
\includegraphics[width=7.5cm]{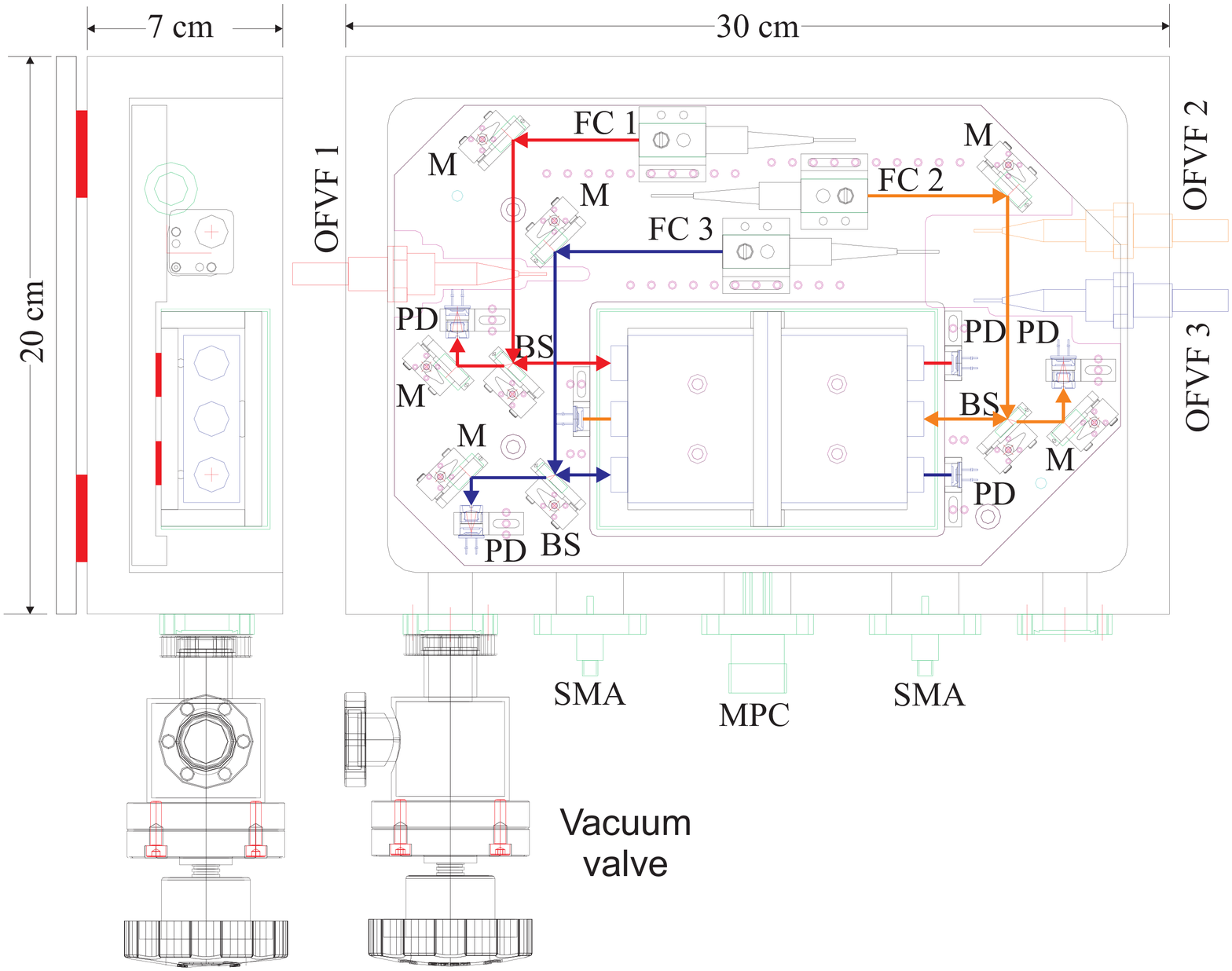}
\caption{Technical layout of the vacuum container and the optics breadboard.\label{fig:tech draw}}
\end{figure}

A schematic of the FSS is shown in Fig. \ref{fig:Principle}. 
The reference for the stabilization is a single, temperature-stabilized
ULE block containing three non-tunable cavities. The block is mounted
in a 30 cm $\times$ 20 cm $\times$ 10 cm UHV chamber which is also
temperature-stabilized. No special effort has been undertaken to minimize
vibration sensitivity, in order to keep the unit simple and compact.
Cavity 3 (fused silica substrates, finesse 1500) stabilizes the 813
nm and 922 nm lasers. Cavity 2 is used for the most demanding stabilization
(689 nm laser), has ULE mirror substrates and a finesse $1\times10^{4}$.
The first cavity is suitable for stabilization of repumper lasers
(679 nm and 707 nm) but this is not currently implemented, since their
passive stability is sufficient. The mirror curvature radii are 0.5
m, and the substrates are AR-coated. Each laser wave passes through
its respective EOPM, and up to two laser waves are combined by a fiber
splitter and coupled to two fundamental modes of a single cavity.
Single-mode fibers with vacuum feed-through deliver the laser light
into the chamber. Fig. \ref{fig:tech draw} shows technical details
of the optics subsystem.

We briefly summarize salient features of the sideband locking technique.
A (typically large) modulation frequency $\omega_{gap}$ bridges the
gap between the target frequency and the reference frequency $\omega_{ref}$.
In addition, a smaller modulation frequency $\Omega_{PDH}$ serves
to produce sidebands for Pound-Drever-Hall (PDH) locking. The drive
voltage $V_{EOM}$ supplied to the EOPM is therefore chosen to be
a PM electric signal, $V_{EOM}=V_{0}\sin\left(\omega_{gap}t+\sin\left(\Omega_{PDH}t\right)\right)$.
The electric field of the laser wave exiting the EOPM is then a ``cascaded''
PM optical wave $E\left(t\right)=E_{L}\exp\left(i\omega_{0}t+i\beta\sin\left(\omega_{gap}t+\alpha\sin\left(\Omega_{PDH}t\right)\right)\right)$.
An expansion of this expression shows that the amplitude of the component
offset from the carrier $\omega_{0}$ by $m\omega_{gap}+n\Omega_{PDH}$,
$m,\: n=0,\:\pm1,\:\pm2,\:....$, is given by sgn$\left(n\right)^{n}$sgn$\left(m\right)^{\left|n\right|+m}J_{\left|n\right|}\left(\alpha\left|m\right|\right)J_{\left|m\right|}\left(\beta\right)$.
Here sgn and $J$ are the sign and Bessel function, respectively.

Consider the ($n=0,\:\pm1$) triplet of sidebands. The two components
$m\omega_{gap}\pm\Omega_{PDH}$ have opposite signs. Therefore, when
the central component $m\omega_{gap}$ interrogates a reference line,
a PDH error signal can be obtained by demodulation of the photodetector
signal at the frequency $\Omega_{PDH}$. The slope of the error signal
for the $m=+1$ triplet (upper 1$^{st}$-order sideband triplet) is
opposite to that of the $m=-1$ triplet (lower 1$^{st}$-order sideband
triplet), since the amplitudes of the sidebands $\omega_{gap}$ differ
by the factor $-1$. Note that there are no PDH sidebands associated
with the carrier (for $m=0$ the amplitudes are zero for any $n$).
If desired, one can make use of the 2$^{nd}$-order sidebands $\pm2\omega_{gap}$
in order to bridge larger gaps $\omega_{ref}-\omega_{0}$, which enhances
the usefulness of the method. A sufficiently strong modulation index
is then required. In our system, we use only the 1$^{st}$-order sidebands.
The slope of the error signal is proportional to $J_{0}\left(\alpha\right)J_{1}\left(\alpha\right)\left[J_{1}\left(\beta\right)\right]^{2}$.
The modulation index $\alpha$ of the electrical signal can be optimized
so as to maximize $J_{0}\left(\alpha\right)J_{1}\left(\alpha\right)$
to 0.34 for $\alpha=1.08$. 

The reference cavities have $\Delta\omega_{f}/2\pi\simeq$~1.48~GHz
free spectral range.

\begin{figure}
\includegraphics[width=7.5cm]{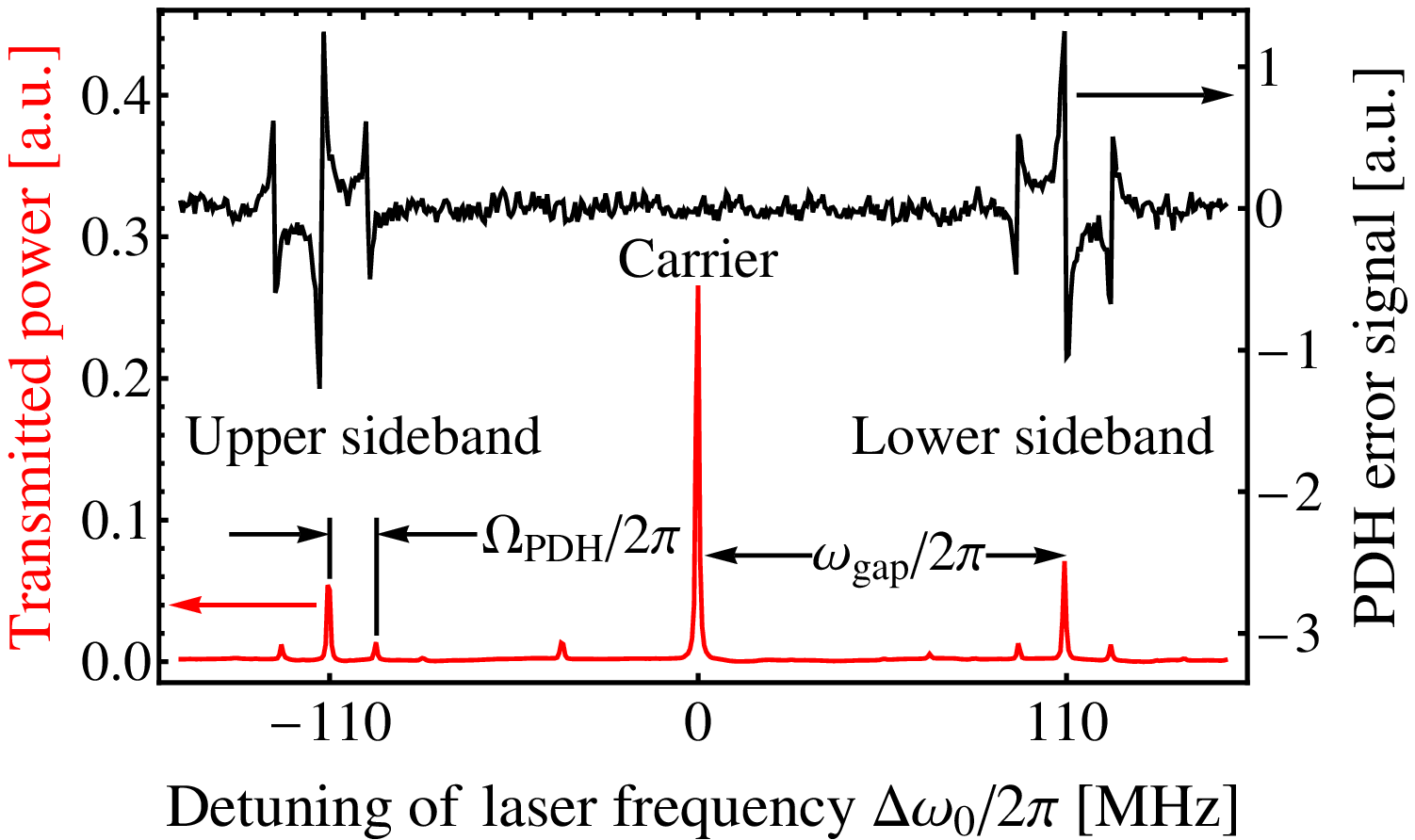}
\caption{Red trace: typical spectrum of the cascaded PM light taken at 689
nm on cavity 2, obtained by scanning the carrier laser frequency $\omega_{0}$
relative to the cavity resonance. The small peak at about $-45$ MHz
is from a higher-order cavity mode. Black trace: PDH error signal
after demodulation at $\Omega_{PDH}/2\pi=12.6$~MHz. Note the absence
of any error signal when the carrier $\omega_{0}$ is in resonance
with the cavity mode (at zero detuning) and the opposite slope of
the error signal for the upper and lower sidebands. Here, $\alpha\simeq0.7,\:\beta\simeq1$.\label{fig:tran-and-PDH}}
\end{figure}

RF signals with frequencies up to nearly this value can be generated
with a low-cost DDS. Our home-built DDS is based on a 400 MHz AD9910
chip (Analog Devices). It provides the required PM signal, whose properties
can be keyed in via front panel or submitted via USB interface. Frequencies
up to 400 MHz are produced directly, while frequencies between 400
and 700 MHz are produced by a AMK-2-13 frequency doubler (Minicircuits)
which is switched on or off by a Mega64L micro-controller (Atmel).
The DDS also provides a separate output $\alpha\sin\left(\Omega_{PDH}\right)$
to the PDH electronics. Finally, the DDS has an external 10 MHz reference
input allowing for highest accuracy and stability. The DDS signal
is amplified by a first amplifier (GALI-24, Mini-Circuits) and further
amplified to approximately 1 W by an external power amplifier (CGD1046HI, NXP).
If gaps $>$ 700 MHz are to be bridged, one can resort to a commercial
microwave synthesizer with PM option.The waveguide EOPM (Jenoptik)
is fiber-coupled and has an electrical bandwidth of 2 GHz. The laser
radiation at the output of the EOPM is coupled into the reference
cavity, and the wave reflected from the cavity is detected as usual
for the PDH technique. The total optical power exiting the collimator
is approximately $80-100$ $\mu$W. The photodetector output is demodulated
with the $\Omega_{PDH}$ reference signal from the DDS, and the error
signal is generated. Fig. \ref{fig:tran-and-PDH} shows the spectrum
of the modulated input light and the error signal.

\begin{figure}
\includegraphics[width=7.5cm]{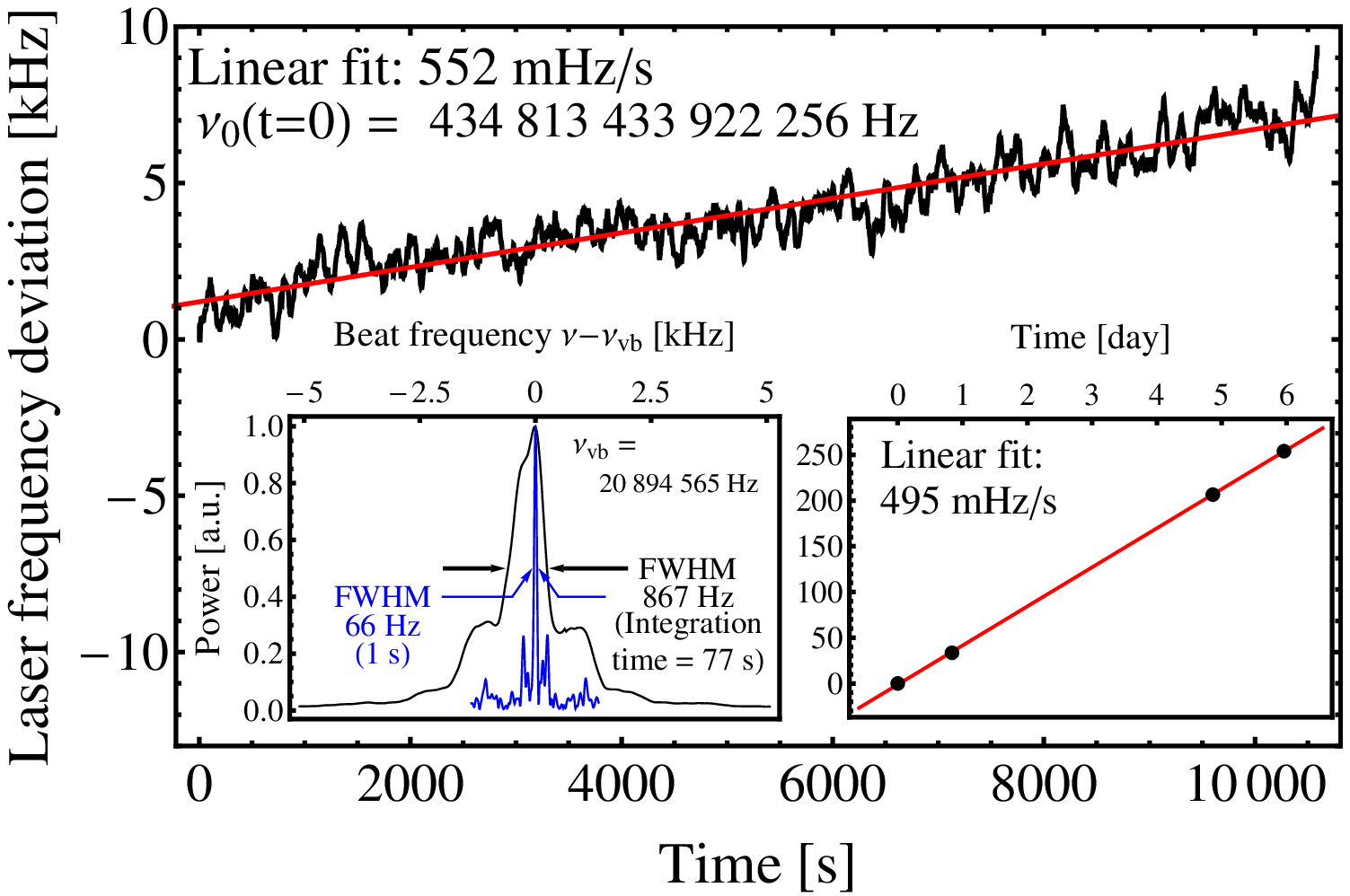}
\caption{Absolute frequency drift of the laser. Left inset: power spectrum of the virtual beat note between
the sideband-stabilized 689 nm laser and an ultra-stable 1156 nm reference
laser. The width of the beat note corresponds to the linewidth of
the 689 nm laser. The sidebands at approximately 1 kHz offset are due to the
acoustic noise of a cooling fan. Right inset: frequency drift over
6 days.\label{fig:drift-and-Allan}}
\end{figure}

The error signal is amplified and filtered by a PID servo that controls
both the diode laser current and the grating position. The choice
of the sign of the servo gain determines whether the upper ($+\omega_{gap}$)
or lower ($-\omega_{gap}$) 1$^{st}$-order sideband is locked to
the cavity resonance.The lasers can be robustly stabilized and remain
in lock for extended periods. The performance of the 689 nm laser
was characterized in detail using the virtual beat technique \cite{Telle2002,Vogt2011}.
By simultaneous provision of a beat of the stabilized laser with a
Ti:Sapphire frequency comb and of an ultra-stable reference laser
(1 Hz linewidth) with the comb, and combination of the two beat notes
and the comb offset frequency into a virtual beat linewidth, the linewidth
can be determined. Additionally, the absolute frequency of the laser
and its drift are determined by optical frequency measurement, with
the comb stabilized to a hydrogen maser. Fig. \ref{fig:drift-and-Allan}
shows the results. A linewidth of approximately 70 Hz on the 1 s time
scale, increasing to 0.9 kHz on the minute-timescale, and a linear
drift of 0.5 Hz/s were achieved. The 922 nm and 813 nm lasers were
characterized by observing the beat of each laser with the comb. The
results show a linewidth less than 1 MHz and a linear drift of 4 Hz/s.

As one aspect of the robustness, the systematic dependence of the
laser frequency on the EOPM drive power has been measured. The sensitivity
is -1.1 kHz/dBm referred to the DDS source output. Taking into account
the measured instability of the amplified EOPM drive signal (approx.
2\% on the time scale of hours), we estimate that the corresponding
frequency uncertainty (relative to the cavity resonance) is of the
order of 100 Hz. It is likely due to residual amplitude modulation
(RAM), since for simplicity and compactness, no countermeasures, such
as  polarization optics , were taken. This uncertainty is sufficiently
low for a 2$^{\text{nd}}$ stage cooling laser of a lattice clock
and also for many other applications.

\begin{figure}
\includegraphics[width=7.5cm]{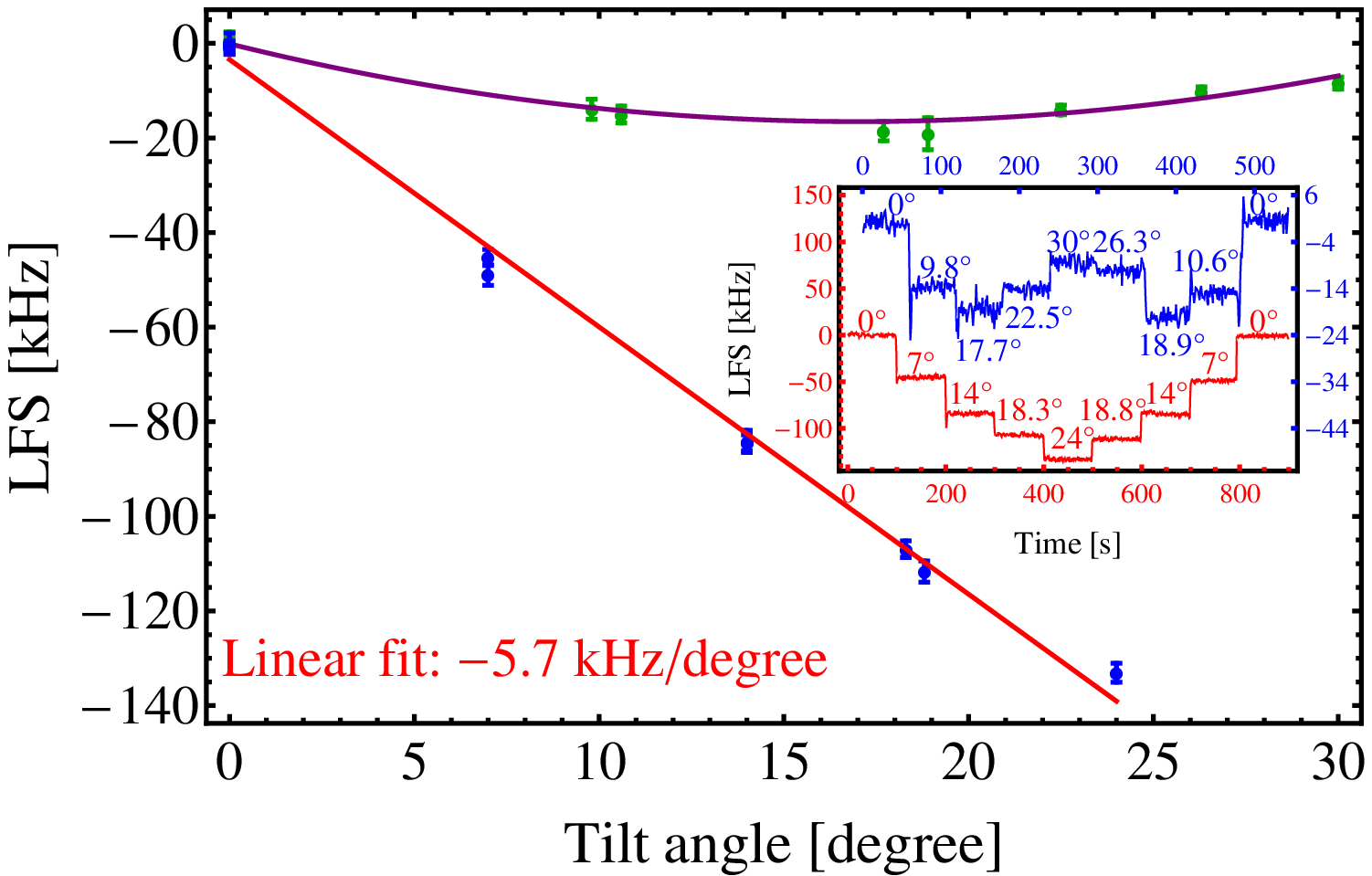}
\caption{Laser frequency shifts (LFS) of the sideband-locked 689 nm laser upon
vacuum chamber tilts. The inset shows the frequency shift versus time
data from which the main plots were generated.\label{fig:Tilt}}
\end{figure}

Another test of the robustness of the FSS consisted in it and observing the shift of
the 689 nm laser locked to it, see Fig. \ref{fig:Tilt}. 
 The frequency shift when the cavity is tilted along its axis is
linear in the tilt angle with sensitivity $-5.7$ kHz/degree. For
tilt around an axis in the optical plane and orthogonal to the cavity
axis, the sensitivity is less than 2 kHz/degree. Considering the tightest
specification for absolute frequency instability, 1~kHz over 1 hour
for the 689 nm laser, an angle stability of 0.1 degree is required
for the device. This is feasible in a stationary application.

\begin{figure}
\includegraphics[width=7.5cm]{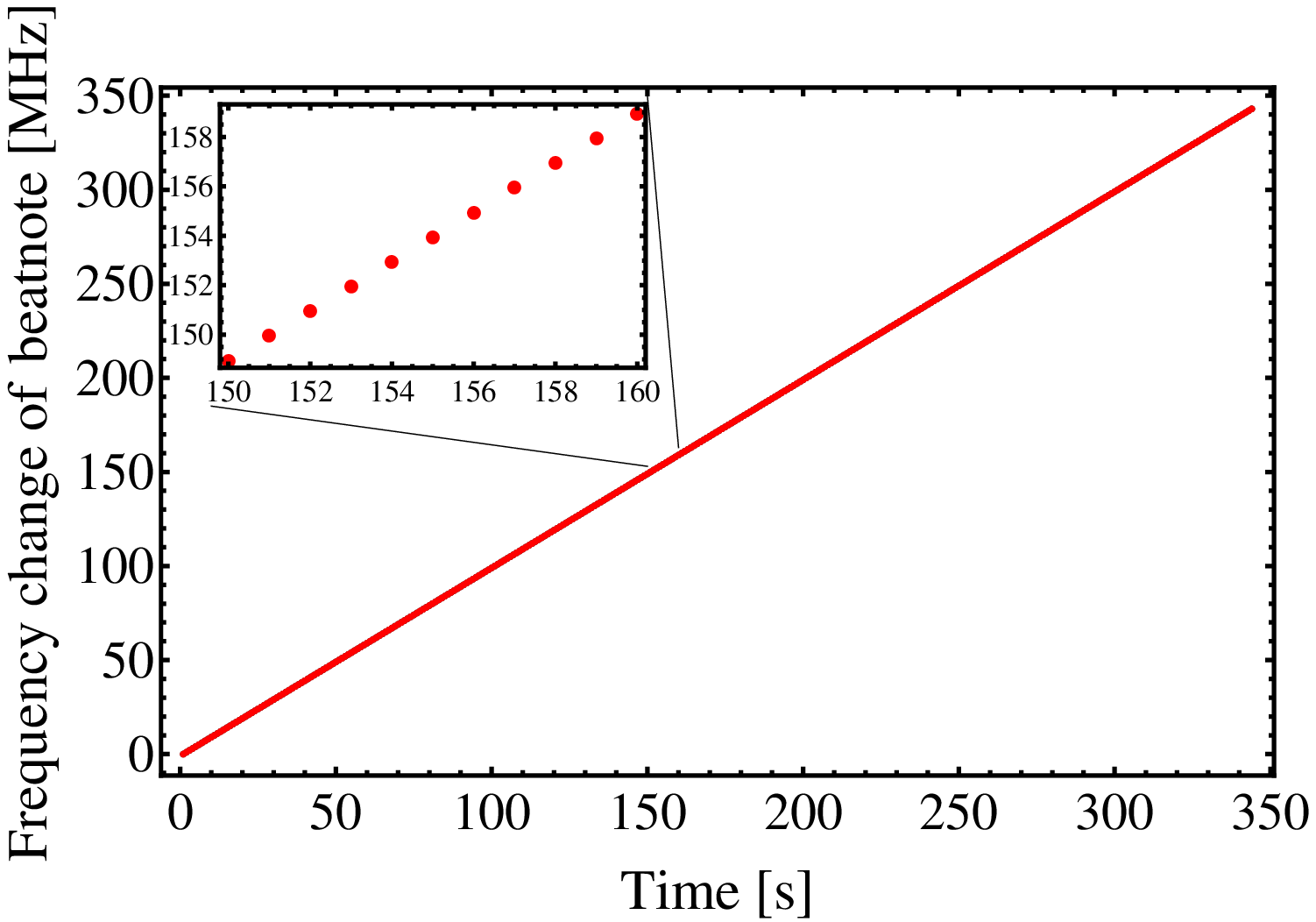}
\caption{Frequency tuning of the carrier frequency of the 689 nm laser while
its sideband was locked to the FSS. A frequency counter recorded the
beat signal with another laser (1 s gate time).\label{fig:sweep}}
\end{figure}

Thanks to the sideband locking technique, an important feature of
the FSS is the freedom to change the carrier frequency $\omega_{0}$
while the laser remains in lock, by changing the gap frequency $\omega_{gap}$.
This is achieved in a robust way by ramping the DDS output frequency
via computer control. The tuning range for $\omega_{0}$ is limited
to $\Delta\omega_{f}/2-\Omega_{PDH}$, because the opposite 1$^{st}$-order
triplet will eventually move over the adjacent resonance and contribute
to the error signal, disturbing the lock. In practice, we have swept
$\omega_{gap}/2\pi$ between 50 MHz and 400 MHz and thus $\omega_{0}/2\pi$
by 350 MHz, see Fig. \ref{fig:sweep}. A tuning speed of 1~MHz/s
is compatible with maintaining lock. The characterization of the sweep
was done by recording the beat between the laser locked to the FSS
and an independent 689 nm laser referenced to the Sr atomic absorption
line.
 Since the modulation index of the EOPM depends on the frequency of
the applied modulation $\omega_{gap}$, the amplitude of sideband
triplet varies with $\omega_{gap}$ even for constant $V_{EOM}$.
Therefore, during the frequency sweep, the gain of the laser current
feedback was reduced in order to prevent the lock from oscillating.
This adjustment was done manually but could in principle be implemented
in the computer control.

The ability to change the gap frequency is very useful, e.g., for
making well-defined frequency scans of the carrier frequency across
a narrow spectral feature. Another use is the compensation of the
frequency drift of the reference cavity: 
the laser frequency drifts can be corrected by modifying periodically
their $\omega_{gap}$ so as to optimize the signals produced in the
excitation of the cold atoms. If the drift of the reference
is known or periodically measured, this information can be used to
digitally change the gap frequency, such that the carrier frequency
$\omega_{0}$ remains stable in time. The FSS is prepared to implement
such a correction using an additional, ultra-stable laser as an absolute
reference, which probes the frequency drift of the ULE block, see
dashed blue subunit in Fig.~\ref{fig:Principle}. In an atomic clock
application, such a laser is naturally available: the clock laser.
Its long-term frequency stability is equal to the atomic clock stability,
since it is steered to the atomic resonance. In our proposed cavity
drift correction concept, the clock laser wave is coupled into its
EOPM and then into cavity 2. The offset frequency $\omega_{gap,ref}$
for the clock laser is set to the resonance of the cavity and an error
signal is generated as for the other lasers. A software analyzes the
error signal and drives it to zero by controlling $\omega_{gap,ref}$~.
The variations of $\omega_{gap,ref}$ with time are monitored and
used to correct digitally the values of $\omega_{gap}$ of the other
locked lasers.

The FSS was shipped from D\"usseldorf to Firenze and integrated into
a \textsuperscript{88}Sr clock apparatus \cite{SOC2}. The 689 nm
laser was stabilized with a home-built PID servo, while the 813 nm
and 922 nm lasers were stabilized using the DIGILOCK 110
(TOPTICA). After locking the laser, the offset frequencies of the
two cooling lasers were tuned so as to optimize the number of trapped
atoms. Approximately, $10^{5}$ atoms were loaded into the optical
lattice via a 1\textsuperscript{st} stage and a 2\textsuperscript{nd}
stage MOT. The atom permanence half-life in the lattice was $\tau=1.52$~s.
Magnetically induced spectroscopy of the clock transition was performed
and a linewidth of 100 Hz was observed.

In summary, we have demonstrated a compact and transportable frequency
stabilization (FSS) system for multiple lasers, in this concrete case
for a Sr lattice optical clock. The FSS uses the resonances of two
medium-finesse ULE cavities inside a single block as frequency references
to stabilize three lasers at 689 nm, 813 nm, and 922 nm. For the laser
with the most stringent requirements a linewidth of 70 Hz and a residual
frequency drift less than 0.5 Hz/s is achieved, a very good result
considering that no efforts have been made to isolate the system from
external vibrations. Additionally, the system is robust as evidenced
by tilting the cavity without loss of lock and with only moderate
frequency change.As proof of functionality of the FSS, the system
was integrated into a transportable Sr atomic clock apparatus. The
initial tests show that, thanks to the FSS, a more stable population
of atoms in the blue MOT, the red MOT, and the lattice are achieved,
as compared to stabilization to an atomic beam spectroscopy signal,
improving the operation of the clock apparatus as a whole. The easy
tunability of the laser frequencies helps to find the best frequencies
for maximum number of atoms stored in the lattice. Besides this particular
application, we believe that similar systems can be used for a variety
of experiments employing cold atoms and molecules.

\textsl{Acknowledgments.} We are grateful to D. Iwaschko, P. Dutkiewicz and U. Sterr (PTB) for support. The research leading to these results has received
funding from the European Union Seventh Framework Programme (FP7/2007-2013
grant agreement 263500).


\end{document}